\documentclass[12pt]{article}

\usepackage{psfig,epsf}

\begin{document}

\begin{flushright}
SWAT/259 \\
\end{flushright}

\vskip 10mm
\centerline{\Large{\bf{Recent Results from (Full) Lattice QCD}}}

\vskip 10mm
\centerline{\bf{C.R. Allton}}

\vskip 10mm
\centerline{\em{Department of Physics, University of Wales Swansea,}}
\centerline{\em{Singleton Park, Swansea SA2 8PP, U.K.}}

\vskip 10mm
\centerline{\tt{E-mail: c.allton@swan.ac.uk}}

\vskip 20mm
An overview of the Lattice technique for studies of the strong
interaction is given.
Recent results from the UKQCD lattice collaboration are
presented. These concentrate on spectral quantities
calculated using full (i.e. unquenched) QCD.
A comparison with quenched results is made.
Novel methods of extracting spectral properties from two-point functions
are described.

\newpage



\section{Introduction}
\label{subsec:intro}

Lattice Gauge Theory has been applied to the study of the strong
interaction in earnest for the last 20 years or so. In that time, it
has grown from an fledgling, optimistic area of research to a
well-developed, mature (and optimistic!) discipline. The reason for
this optimism is that all the approximations involved in the technique
are systematically improvable. This means that, given enough
time on a computer powerful enough, we will have predictions for the
QCD bound-state spectrum (for example) with arbitrary accuracy. Other
approaches of studying the strong interaction, while they may have
their own advantages, cannot make this claim.

In this talk I will begin by overviewing the lattice method for
obtaining hadronic spectral quantities via the calculation of $n-$point
functions. I will outline the caveats that exist with current lattice
simulations as well as detailing some of the successes of the approach.
The vexed question of performing ``full'' QCD calculations (i.e.
without the quenched approximation) is discussed and I will explain why
these simulations are even more difficult than they at first seemed. In
Sec. \ref{sec:ukqcd}, I will detail some recent results from the UKQCD
collaboration, focusing on our recent full QCD work. Sec. \ref{sec:spf}
outlines a new and promising approach of obtaining physical results
from the lattice by using the spectral function representation. This is
an exciting area of research, and, if it reaches its potential,
promises to become a standard approach in the future. I then summarise
the main points raised in this talk in Sec. \ref{sec:fini}.



\section{Overview of Lattice Gauge Theory}\label{sec:overview}

\subsection{The Method}\label{subsec:method}

I attempt here to describe the method normally used in lattice
calculations of the hadronic spectrum of QCD. There are many
excellent reviews of this topic which cover the approach
in more detail.\cite{reviews}

The conventional lattice approach to QCD spectrum calculations is performed
in the Euclidean path integral framework. It requires the
calculation of $n$-point correlation functions, $G_n(t)$, of hadronic
interpolating operators, $J$, in a background sea of glue (and sea quarks in
the case of full, i.e. ``unquenched'' calculations).
A Monte Carlo approach is used to generate these background
configurations with the appropriate Boltzmann factor $e^{-{\cal S}}$
where ${\cal S}$ is the Euclidean action. There is an obvious analogy
between this Lagrangian approach to Lattice Gauge Theory, and
statistical mechanics: clearly the lattice path integral corresponds to
the partition function of statistical mechanics.

Note that in lattice simulations there is freedom to choose different
quark fields in the interpolating operator $J$ compared to those in
the Lagrangian. Hence we are able to distinguish between ``valence'' and
``sea'' quark masses, $m^{val}$ and $m^{sea}$.

Because the QCD Lagrangian contains fermionic fields that appear
quadratically, they can be integrated out analytically giving the
usual determinant factor, $det( D\!\!\!\!/ + m )$. Including this
factor is a technical headache because it involves a huge increase in
computational requirements. The usual way around this is to invoke the
quenched approximation where two things are done: (i) the fermionic
determinant is replaced by unity; and (ii) the gauge coupling, $g$, is
rescaled so that the physical predictions for a particular test
quantity (like the rho mass, for example) is in agreement with its
experimental value. The extent to which the quenched approximation
reproduces physical predictions for other quantities is a measure of
its success. It is a remarkable fact that for many spectral
quantities, the quenched predictions agree with the experimental
(i.e. the ``full'' QCD) values to within 10\% (see \cite{cppacs}).

Despite the success of the quenched approximation, it is obviously
essential to perform full QCD calculations in order to study the real
world. Furthermore, for some quantities (such as the deconfinement
temperate\cite{karsch}) the quenched approximation is poor, and for others
(such as the $\eta'$ mass) the quenched approximation fails completely.

In the calculation of the two-point function, $G_2(t)$, if the exact
operator, $J^{exact}$ for the hadron in question was used, then
$G_2(t)$ would contain information on that hadron and no others.
However, since $J^{exact}$ is not known, $G_2(t)$ receives
contributions from all hadronic states which have non-zero overlap with
$J$. It is straightforward to show that, in this case, $G_2(t)$ has the
following form
\[
G_2(t) = \sum_{i} Z_i e^{-M_i t}
\]
where the sum is over the hadronic states $i$, and $M_i$ and $Z_i$ are the
corresponding hadronic mass and overlap. Note that since the
calculation is performed in Euclidean space-time, the excited states
are exponentially suppressed with respect to the fundamental state
(i.e. the exponentials have real arguments). This means that $G_2(t)$
asymptotes to the two-point function of the ground state hadron
as $t \rightarrow \infty$. Obviously the parameters of the
ground state, in particular the mass $M \equiv M_0$, can be extracted by
fitting $G_2(t)$ to an exponential, $e^{-M t}$, for $t$ sufficiently large.



\subsection{Caveats}\label{subsec:caveats}

In this sub-section I explain the caveats that one must apply to any
calculation using the method described above.
The main point to make is that the calculations of the hadronic
properties are performed with input parameter values which are
{\em not} those of the real world.
Specifically, these parameters are 
\begin{itemize}
\item valence quark mass(es), $m^{val}$, (typically $m^{val} 
\;\raisebox{-.5ex}{\rlap{$\sim$}} \raisebox{.5ex}{$>$}\;\frac{1}{2}m^{strange}$);
\item sea quark mass(es), $m^{sea}$, (typically $m^{sea}
\;\raisebox{-.5ex}{\rlap{$\sim$}} \raisebox{.5ex}{$>$}\;m^{strange}$);
\item lattice volume, $V$, (typically $V 
\;\raisebox{-.5ex}{\rlap{$\sim$}} \raisebox{.5ex}{$<$}\;(2\;fm)^3$);
\item lattice spacing, $a$, or, through dimensional transmutation, $g_0$,
(typically $a
\;\raisebox{-.5ex}{\rlap{$\sim$}} \raisebox{.5ex}{$>$}\;.05\;fm$);
\item number of dynamical fermion flavours, $N_f$.
Quenching corresponds to $N_f = 0$. (Typically, $N_f = 0$ or $2$.)
\end{itemize}
Therefore the mass, $M$, which was obtained using the procedure above is
not the mass of the real world hadron, but the mass of the
corresponding hadron in a world where the quark masses are the same
as those input into the lattice calculation, the volume is the finite
volume used in the simulation etc.
So, strictly speaking, $M$ is a function of the above 5 input parameters.
The final prediction of the real world value, $M^{expt}$, should be
obtained by the following extrapolations:

\begin{eqnarray}\nonumber
\surd\surd\surd&\;\;\;\;	m^{val} &\rightarrow \mbox{few}\;\; MeV \\\nonumber
\surd\surd&\;\;\;\;		m^{sea} &\rightarrow \mbox{few}\;\; MeV \\\nonumber
\surd\surd\surd&\;\;\;\;	V &\rightarrow \infty \\\label{eq:extraps}
\surd\surd\surd&\;\;\;\;	a &\rightarrow 0 \\\nonumber
\surd&\;\;\;\;			N_f &\rightarrow \mbox{``}2 \frac{1}{2}\mbox{''}
\;\;\;\mbox{i.e. two light flavours for }u,d\mbox{ and one heavier for }s
\end{eqnarray}
Note that the limit $m^{sea} = \infty$ corresponds to $N_f = 0$ so
the $m^{sea}$ and $N_f$ extrapolations are not independent.  While these
extrapolations muddy the water significantly, many of them are
theoretically well-understood and numerically under control. The
number of $\surd$'s which appear next to the extrapolations is an
indication of how well the extrapolation is under control.

\begin{figure}[htb]
\vspace*{5mm}
\epsfxsize=0.65\textwidth                                         
\epsfbox[-50 60 450 560]{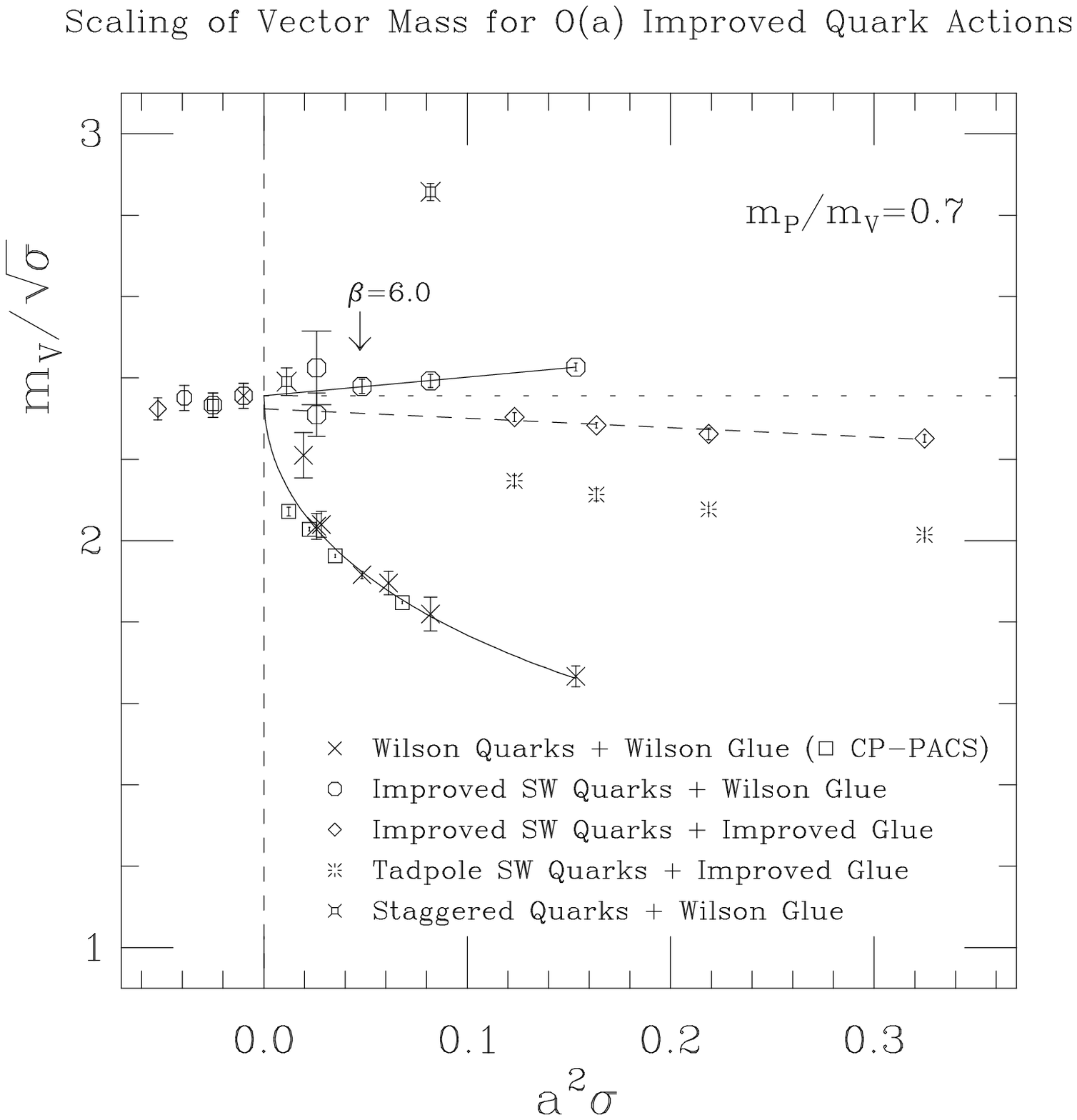}
\vspace*{-10mm}
\caption{Comparison of scaling violations in the vector meson mass,     
computed in quenched QCD using the standard Wilson action and
several different ${\cal O}(a)$-improved actions from \cite{ordera}.}
\label{fig:a}
\end{figure}

As an example of the quality of the extrapolations,
Fig. \ref{fig:a} shows an extrapolation of the nucleon and vector
meson masses as a function of the lattice spacing, $a$, taken from
\cite{ordera}.  In these plots, data points obtained with both the
Wilson and various improved lattice actions (designed to have
discretisation errors smaller than ${\cal O}(a)$) are shown.
Fits to the relevant functional forms are included in the figure
and the symbols on the left of the plot are the continuum extrapolations
(which clearly agreement with each other).

\subsection{Successes}\label{subsec:Success}

\begin{figure}[htb]
\vspace*{20mm}
\epsfxsize=0.6\textwidth
\epsfbox[-100 60 400 560]{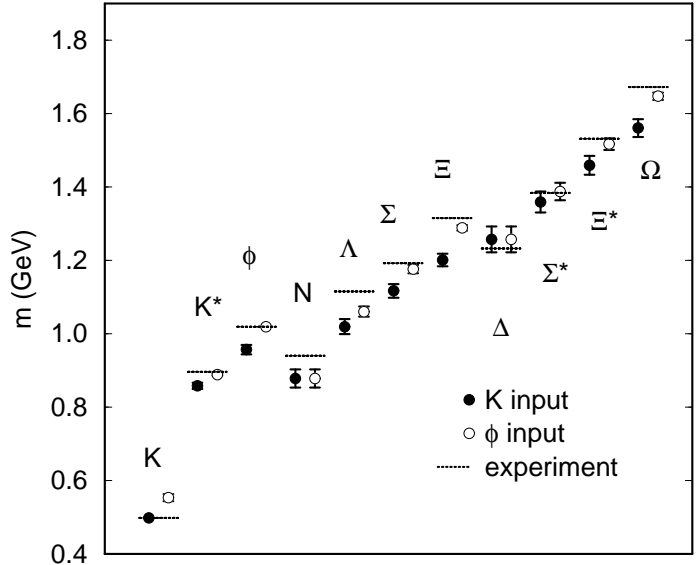}
\vspace*{-30mm}
\caption{CP-PACS results for the light hadron spectrum in quenched QCD
obtained using the Wilson quark action.\cite{cppacs}}
\label{fig:cppacs}
\end{figure}

The caveats listed in the previous sub-section do not hinder the success
of the Lattice technique as a means of obtaining accurate predictions from
the strong interaction. To give an example of the current status of
lattice calculations, Fig. \ref{fig:cppacs} shows the hadronic spectrum
obtained by the CP-PACS collaboration using the {\em quenched}
approximation.\cite{cppacs}
There are two features to note.
The error bars in the predictions are tiny
($\;\raisebox{-.5ex}{\rlap{$\sim$}} \raisebox{.5ex}{$<$}\;$3\%) - which is a
clear measure of the success of the lattice technique.
Secondly, there is a small, but statistically significant discrepancy
between the lattice predictions and the experimental numbers - which
is a signal that unquenching is required in order to make further
progress.

While the quenched lattice calculations have clearly matured with precision
estimates of many quantities of only a few percent, serious calculations
involving full QCD have only recently begun. Typically the current errors
in these calculations are several times that of equivalent quenched results.




\section{Recent Dynamical Results from the UKQCD Collaboration}\label{sec:ukqcd}

In this section I review some of the recent results from the UKQCD
Collaboration's dynamical simulations.\cite{ukqcd1,ukqcd2}


\subsection{Dependency on Sea Quark Mass}

\begin{figure}[htb]                                                  
\epsfxsize=0.65\textwidth                                         
\epsfbox[-50 60 450 560]{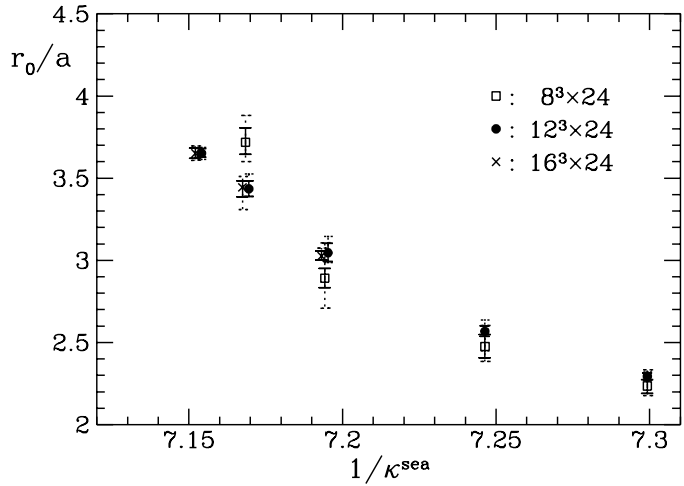}
\vspace*{-30mm}
\caption{The inverse lattice spacing plotted against quark mass
(expressed in terms of $1/\kappa^{sea}$) for different lattice sizes taken
from \cite{ukqcd1}.
The chiral limit is approximately at the left margin of the figure.}
\label{fig:r0a}
\end{figure}

As was outlined in Sec. \ref{subsec:caveats} all lattice predictions
are functions of $V$, $a$, $N_f$ etc. In this section, the functional
dependency on $m^{sea}$ is discussed. Fig. \ref{fig:r0a} (taken from
\cite{ukqcd1}) shows how the lattice spacing, obtained from the
hadronic length scale\cite{sommer}, $r_0 \approx 0.5 fm$,
depends strongly on the sea quark mass, $m^{sea}$ (here expressed
in terms of the hopping parameter $\kappa^{sea}$). These
calculations were performed at fixed bare coupling, $g_0$.

This effect has important consequences. Simulations at a fixed bare
coupling, $g_0$, and with several values of $m^{sea}$, correspond to
different physical volumes, and, furthermore, different points on the
continuum extrapolation $a\rightarrow 0$. Thus finite volume and
${\cal O}(a)$ systematics are mixed in this case.

UKQCD have performed two types of calculation. The first was at fixed
$g_0$ for various values of $m^{sea}$ in order to calibrate this
effect\cite{ukqcd1}. We then performed more sophisticated simulations
at several values of the parameter pairs $(g_0, m^{sea})$ which were
chosen in order to maintain {\em fixed} lattice spacing $a \approx
0.11 fm$ (and therefore also fixed volume)\cite{ukqcd2}. This second
calculation utilised the ``matching'' technology of \cite{alanjim}.
In both these calculations, an improved action was used in order to
reduce the effect of ${\cal O}(a)$ errors.



\subsection{Results}

There is not space to present full details of UKQCD's recent
unquenched calculations. I discuss the results of only two quantities,
and refer the reader to the original papers for full
details.\cite{ukqcd1,ukqcd2}

\begin{figure}[htb]
\epsfxsize=0.6\textwidth
\vspace*{10mm}
\epsfbox[-100 60 400 560]{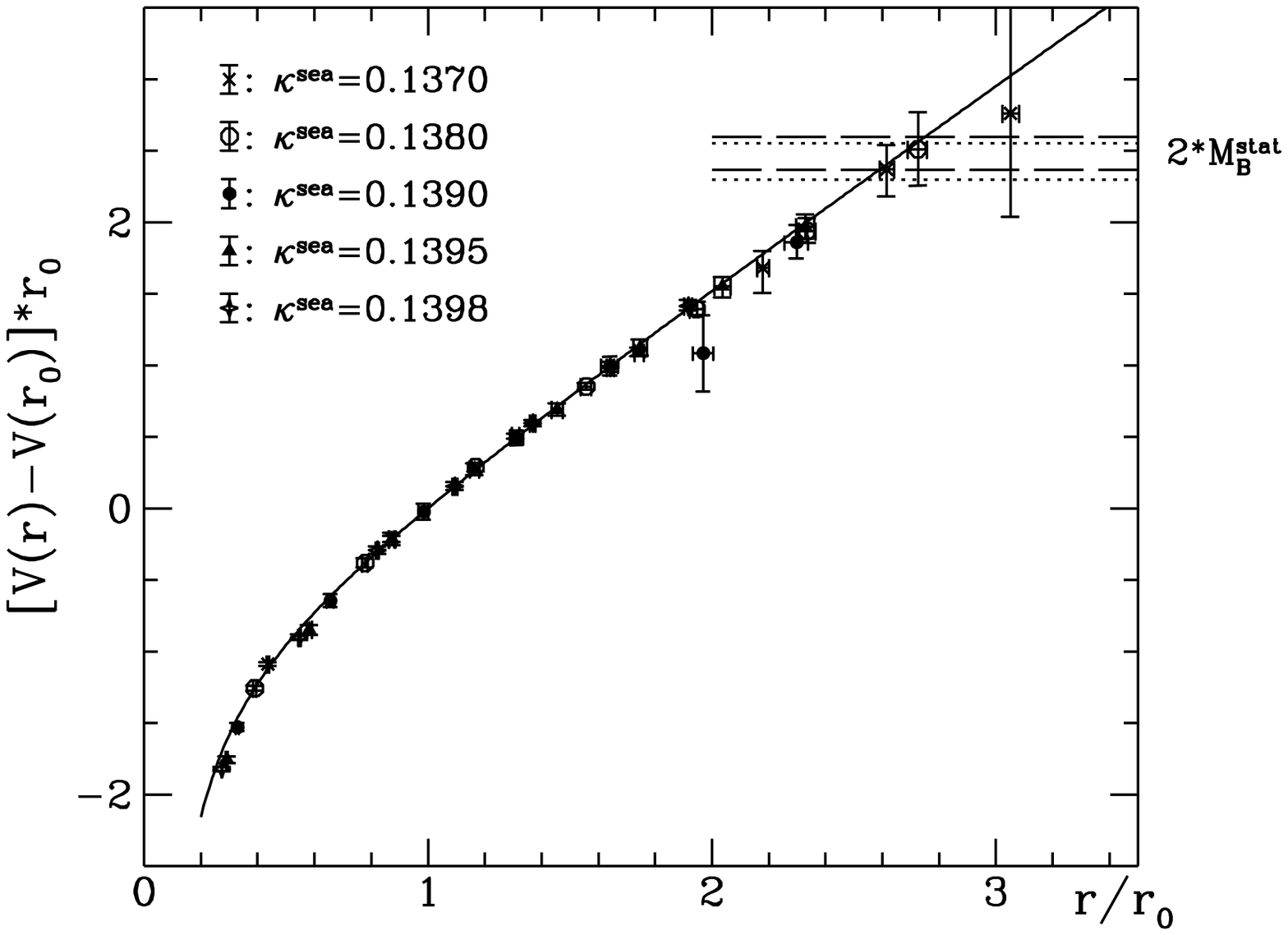}
\vspace*{-30mm}
\caption{The scaled and normalised potential as a function of distance
(from \cite{ukqcd1}).}
\label{fig:stat_pot}
\end{figure}

One of the benchmark quantities of lattice calculations is
the static quark potential. Fig. \ref{fig:stat_pot} shows UKQCD's
result for this quantity in units of $r_0$ from \cite{ukqcd1}. It can
be noted that there is little immediate dependency on $m^{sea}$ in this
quantity.\footnote{A closer look however at the data points close to
the origin shows a systematic effect which can be interpreted as
different runnings of the coupling as function of $m^{sea}$.}

\begin{figure}[htb]                                                  
\epsfxsize=0.65\textwidth                                         
\epsfbox[-50 60 450 560]{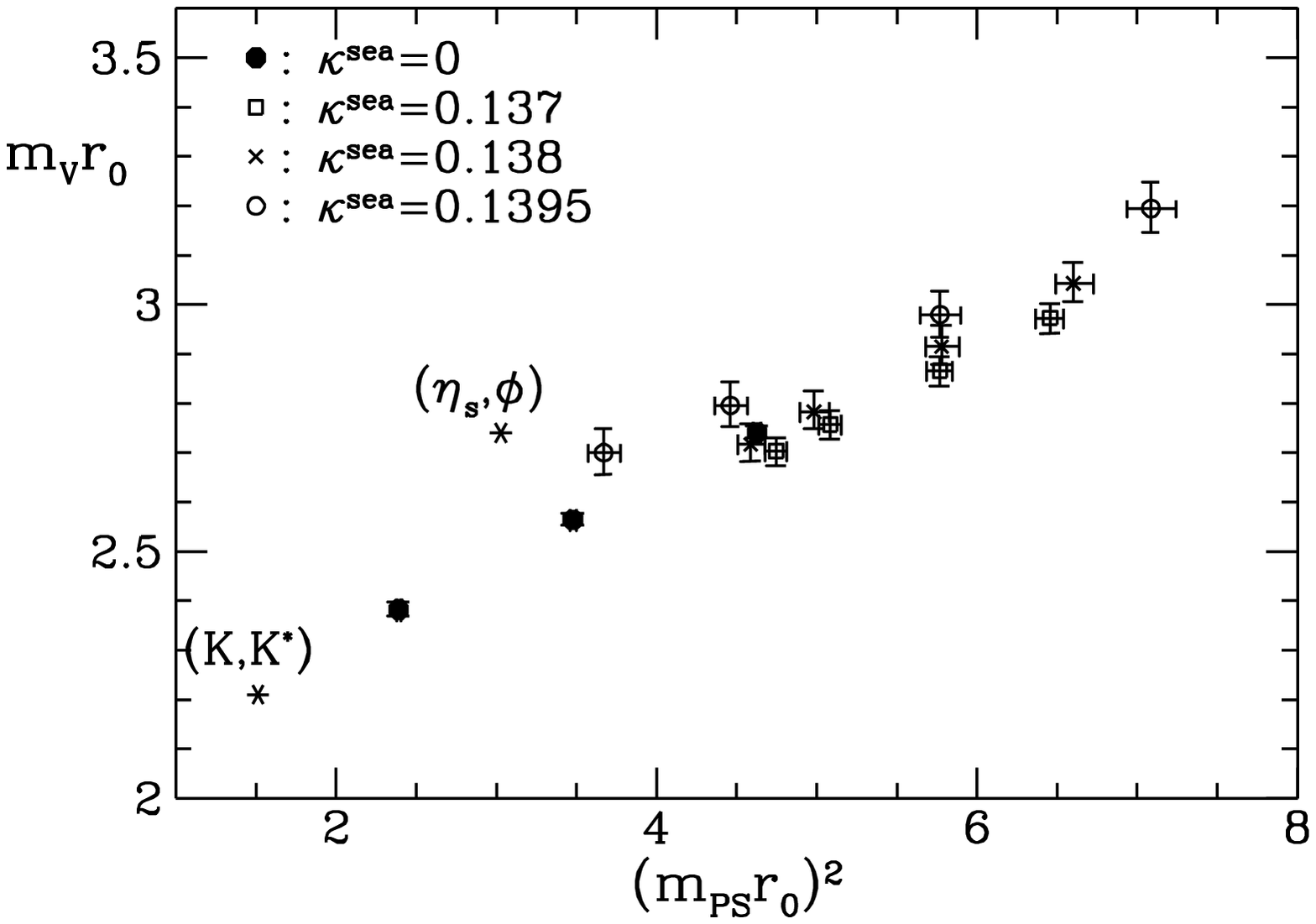}
\vspace*{-30mm}
\caption{Vector mass plotted against the pseudoscalar mass
for several sets of different sea quark masses from \cite{ukqcd1}
The experimental points are denoted by $\ast$
and the quenched results are labelled $\kappa^{sea}=0$.}
\label{fig:mvr0}
\end{figure}

Fig. \ref{fig:mvr0} shows the vector meson mass, $M_V$, plotted
against the pseudoscalar mass squared, $M_{PS}^2$ from \cite{ukqcd1}.
Again, these quantities are expressed in units of $r_0$. Quenched data
which corresponds to the same lattice spacing as the unquenched
simulations are included as a comparison.  The experimental points
corresponding to the strange mesons are also plotted.

In summary, the results of \cite{ukqcd1,ukqcd2} indicate that the
effects of unquenching are small for these values of $m^{sea}
\;\raisebox{-.5ex}{\rlap{$\sim$}} \raisebox{.5ex}{$>$}\;m^{strange}$.
This motivates the need to move to more physical
values of $m^{sea} \approx \mbox{few } GeV$ in the future.




\section{Lattice Spectral Functions}\label{sec:spf}

As outlined in Sec. \ref{subsec:method}, the conventional method of
determining ground state properties from lattice simulations is by
fitting exponentials to the tails of $n$-point functions.  There have
been several attempts at developing other strategies for uncovering
spectral quantities from lattice data.\cite{derek,simon,def,orange,dbjc}
These all revolve around the spectral
function (SF) $\rho(s)$ which can be defined through
\begin{equation}\label{eq:sf}
G_2(t) = \int_0^\infty K(t,s) \rho(s) ds,
\end{equation}
where $K(t,s)$ is the kernel function - typically just $e^{-st}$ for
this work. The SF contains much richer information on the channel
being considered than just the ground state parameters. It also has
the advantage that theoretical input can be used to guide its form for
large values of $s$ where perturbation theory is valid.  In fact, this
was the approach taken by \cite{derek,orange} who used (continuum)
perturbation theory to derive a functional form for
$\rho^{PT}(s)$. This $\rho^{PT}(s)$ was then used above a certain
threshold of energy $s_0$ and a $\delta$ function used for the ground state
following the approach of QCD Sum Rules.

A very new and promising technique takes the marriage of lattice data
and spectral functions one step further.\cite{mem}  This approach
uses the lattice data itself to determine the SF by inverting
eq.(\ref{eq:sf}). This is a very numerically technical approach; it is
in fact an ``ill-posed'' problem - $G_2(t)$ is known only at a small
number of discrete values of $t$, whereas the aim is to determine
$\rho(s)$ for a large number of values of $s$ (ideally for a
continuous range in $s$).  In \cite{mem} a ``maximum entropy method''
is employed to overcome these hurdles.



\section{Conclusion}\label{sec:fini}

I have given a brief overview of the current state of play of lattice
gauge theory calculations of the hadronic spectrum.  The problems of
extrapolating the input parameters of any lattice simulation to
their physical value is emphasised.  In particular I have shown that
interpreting results from naive unquenched calculations can prove
difficult due to the dependency of the lattice spacing on $m^{sea}$.
A summary has been given of recent unquenched results from the UKQCD
collaboration for the static quark potential and vector meson mass.
Finally the new and interesting method of using spectral functions
in the analysis of lattice data is discussed.



\section*{Acknowledgements}
I am grateful to my colleagues in the UKQCD collaboration especially
Joyce Garden, Balint Joo, Alan Irving and Ken Bowler.
I acknowledge the support of the Particle Physics \&
Astronomy Research Council and the Royal Society.



\end{document}